\documentclass{aa}

\usepackage{graphics}
\usepackage{epsfig}
\usepackage{latexsym}
\newcommand{\be} {\begin{equation}}
\newcommand{\ee} {\end{equation}}
\begin{document}
\hyphenation{brems-strah-lung}
\title{On hadronic beam models for quasars and microquasars}

\author{Diego F. Torres$^{1}$, \& Anita Reimer$^2$}

\institute{
ICREA \& Institut de Ci\`encies de l'Espai (IEEC-CSIC), Campus UAB, Facultat de Ci\`encies, Torre C5-parell, 2a planta, 08193 Barcelona, Spain
\and
Institut f\"ur Theoretische Physik, and Institut f\"ur Astro- und Teilchenphysik, Leopold-Franzens-Universit\"at Innsbruck,  Technikerstr. 25, 6020 Innsbruck, Austria 
}

\offprints{Diego F. Torres. E-mail: dtorres@ieec.uab.es}
\mail{}

\date{ }

\date{Received  / Accepted }

\titlerunning{Hadronic beams}
\authorrunning{Torres \& Reimer}

  \abstract
   {Most of the hadronic jet models for quasars (QSOs) and microquasars (MQs) found in literature represent beams of particles (e.g. protons). These particles interact with the matter in the stellar wind of the companion star in the system or with crossing clouds, generating $\gamma$-rays via proton-proton processes.} 
  {Our aim is to derive the particle distribution in the jet as seen by the observer, so that proper computation of the $\gamma$-ray and neutrino yields can be done.
 } 
{We use relativistic invariants  to obtain the transformed expressions in the case of a power-law and power-law with a cutoff particle distribution in the beam. We compare with previous expressions used earlier in the literature.}
{We show that formerly used expressions for the particle distributions in the beam as seen by the observer are in error, differences being strongly dependent on the viewing angle. For example, for $\Gamma =10$  ($\Gamma$ { is the Lorentz factor of the blob)} and angles larger than $\sim 20^o$,
the earlier-used calculation entails an over-prediction (order of magnitude or more) 
of the proton spectra for $E>\Gamma
mc^2$, whereas it always over-predicts (two orders of magnitude) the proton spectrum 
at lower energies, disregarding the viewing angle. }
{ All the results for photon and neutrino fluxes in
hadronic models in beams that have made use of the earlier calculation are affected. Given that correct $\gamma$-ray fluxes will be in almost any case significantly diminished in comparison 
with published results, and that the time of observations in Cherenkov facilities 
grows with the square of the flux-reduction factor in a statistically
limited result, the possibility of observing hadronic beams is  undermined. }
{} 

   \maketitle

\section{Introduction}

The study of the possible high-energy radiation from QSOs and MQs lacks 
certainty in one central aspect, the particle composition of the jets. Is 
the radiation emitted dominantly produced via inverse Compton (self-synchrotron and/or with external fields) or via proton-proton/proton-photon 
interactions leading to subsequent meson decay? In the latter case, hadronic jet models have been used in order to assess this possibility. 
Most of these models represent beams of particles (e.g. protons) linearly propagating in a direction normal to an accretion disc, interacting 
with the matter in the stellar wind of the companion star in the case of MQs, or with matter in clouds, in the case of quasars. The protons in the beams are usually assumed to be distributed with a power-law
which may be cut at high-energies, as would be the case if they are accelerated in Fermi processes within the jet and subject to losses.
It is then crucial to have a correct description of the proton distribution as seen by the observer, since the emissivity of 
$\gamma$-ray and neutrino produced by charged and neutral pions make use of cross sections known in that frame. In this paper we present in detail the derivation of 
a beam particle distribution as seen by the observer and compare it with the earlier-used expression, finding significant differences.
Consequences are discussed.

\section{Derivation of the particle distribution}

In what follows, primed quantities
refer to quantities that reside in the jet frame, whereas unprimed quantities are used for those in the observer frame.
We consider the Lorentz transformations  
for energies and momenta 
\begin{eqnarray}
E &=& \Gamma (E'+\beta c p'_\parallel), \hspace{0.2cm}
E' = \Gamma (E-\beta c p_\parallel), \nonumber \\
p_\parallel &=& \Gamma (p'_\parallel + \beta E'/c),\hspace{0.2cm}
p'_\parallel = \Gamma (p_\parallel-\beta E/c),
\label{2}
\end{eqnarray}
where $
p_\parallel = p \cos(\theta) $ and $
p_\perp =p \sin(\theta) $ are the parallel and
perpendicular momentum, respectively.
For the relativistic case, $m c \ll p$ and $m c^2\ll E$, the transformation of the energy simplifies to
$
E = D E' , $  with
$D = { 1} / [\Gamma (1-\beta \cos(\theta))].$ 


The elementary invariants under Lorentz transformations are (see, e.g., Dermer \& Menon 2010) 
the invariant four-volume:
$
d{\bf x} dt = dV dt,
$
the invariant phase-space element:
$
{d{\bf p} }/{E} ={p^2 dp d\Omega}/{E} \rightarrow \epsilon d\epsilon d\Omega,
$
with the final expression applying to photons and extremely relativistic particles,
and the invariant phase volume:
$
d\nu = d{\bf p} d{\bf x},
$
where bold-faces stand for three-dimensional space magnitudes. Because the number $N$ of particles or photons is invariant, one has that 
\be
\frac{dN}{d\nu} = \frac {1}{p^2} \frac{dN}{dV dp d\Omega} 
\label{a1}
\ee
is also invariant. For photons and extremely relativistic particles the last invariant becomes
\be
 \frac {1}{\epsilon^2} \frac{dN}{dV d\epsilon d\Omega}. 
 \label{aa1}
\ee
Use of the equality 
$
E^2 = c^2 p^2+m^2 c^4, 
$
allows to write 
$
p^2 dp = (p E / c^2) dE.
$
Use of the latter in Eq. (\ref{a1}) implies that the expression 
\be
\frac {1}{p E} \frac{dN}{dV dE d\Omega} 
\label{a2}
\ee
is invariant under Lorentz transformations. Again, for photons and extremely relativistic particles 
$pc=E$, ($=\epsilon$, in the notation above for this case) and the last invariant becomes expression (\ref{aa1}).
We define the differential number density of particles as 
$
n(E,\Omega) \equiv {dN} / [{dV dE d\Omega}] ,
$
i.e., the differential number of photons or relativistic particles with dimensionless energy between $E$ and $E + dE$ 
that are directed into differential solid angle interval $d\Omega$ in the direction ${\Omega}$ of some physical volume $dV$.
Because of the invariant (\ref{a2}) and the definition of $n(E,\Omega)$ we can write
that 
\be
n(E,\Omega) = n'(E',\Omega') \left( \frac p{p'} \right)  \left( \frac E{E'} \right),
\label{gen-inv}
\ee
which is valid in general,  and
\be
n(E,\Omega) = n'(E',\Omega') \left(\frac {E^2}{E'^2}\right),
\label{rel-inv}
\ee
which is valid for photons and extremely relativistic particles.
Eq. (\ref{gen-inv}) just says that the expression for the distribution function
$f({\bf r,p})$ appearing in kinetic theory of gases, such that the product $f({\bf r,p}) d{\bf p} dV$ is the number of particles lying in the
a given volume element $dV$ and having momenta in definite intervals $d{\bf p}$,
is Lorentz-invariant (see e.g., Eq. 10.5 of Landau \& Lifshitz 1987). Note that 
\be
f({\bf r,p}) = \frac{dN}{dV\,d{\bf p}} = \frac{dN }{ dV\, p^2\, dp \, d\Omega} = \frac{ dN }{ (p E / c^2) \, dV \, dE \, d\Omega},
\ee
and the invariance of 
$f({\bf r,p})$ is that expressed by Eq. (\ref{a2}).

\subsection{The extremely relativistic case}

In this case the particle density in the jet frame, if it is a power-law defined as
\be
n'(E',\Omega') = (A/4\pi)  E'^{-\alpha},       
\label{nn'2}                       
\ee
transforms into the observer frame, by virtue of the previous formulae (\ref{rel-inv}, and \ref{nn'2}), as,
\begin{eqnarray} 
\label{us}
n(E,\Omega) & = & (E/E')^2 n'(E',\Omega')  =  (A/4\pi) D^2 (E/D)^{-\alpha} \\ \nonumber
           &= & (A/4\pi)  E^{-\alpha}  D^{2+\alpha}.
\end{eqnarray}  

 \begin{figure}
\centering
\vspace{0.4cm}
\hspace{0.5cm} \includegraphics[angle=-90, scale=0.3]{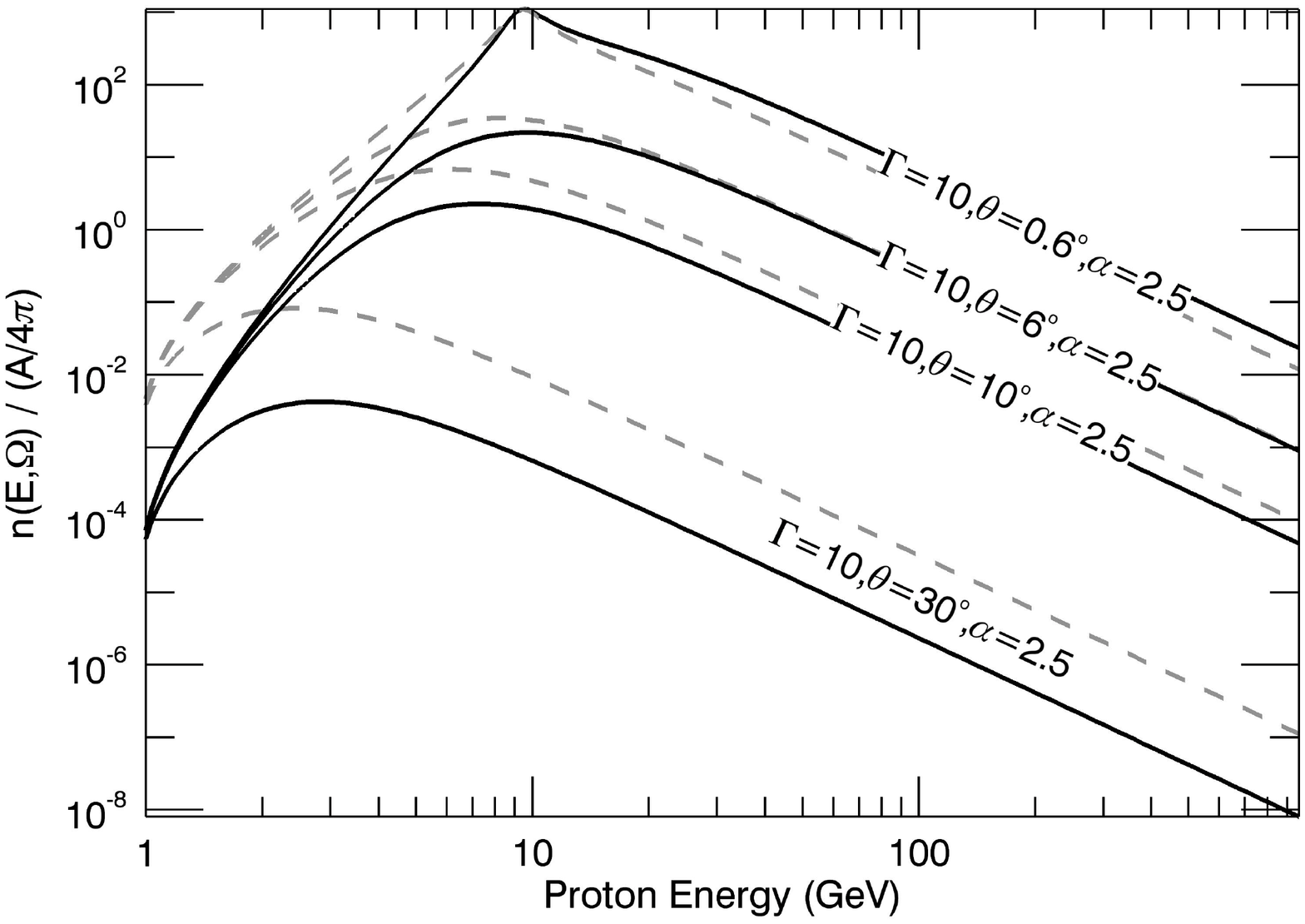} 
     \caption{A power-law spectrum of protons in a beam as seen by the observer for different values of viewing angles, and an example of Lorentz factor and slope (solid lines).
The grey-dashed lines stand for the earlier derivation results for the same parameters (see \S \ref{CT}).}
         \label{n-fig}
\end{figure}

\subsection{The general case}

Using Equations (\ref{2}) and the one for $E^2$,
it is possible to find that
\begin{eqnarray} 
E/E' & = & E/[\Gamma (E-\beta  c p \cos(\theta))]  \nonumber \\
& = & \frac{1}{  \Gamma \left(1-\beta \cos(\theta) \sqrt{1-m^2 c^4/E^2} \right) }.
\label{ee'}
\end{eqnarray} 
In addition, using the Lorentz transformation for momenta (Eq. \ref{2} and the fact that $p_\perp =p \sin(\theta) =p_\perp ' $), one has
\begin{eqnarray} 
p'^2 &=& p^2 \sin^2 (\theta) + \Gamma^2 ( p \cos(\theta) - \beta E / c)^2 )  \nonumber \\
          &=& p^2 \left[ \sin^2 (\theta) + \Gamma^2 (  \cos(\theta) - \beta E / \sqrt {E^2 - m^2c^4} )^2 \right].
          \label{pp'}
\end{eqnarray} 
Using the latter expression to get $p/p'$ and inserting this into Eq. (\ref{gen-inv}),   together with
Eq. (\ref{nn'2})   and Eq. (\ref{ee'}) we get, 
\begin{eqnarray} 
\label{final}
n(E,\Omega)  &=&  \frac{ A}{4\pi} \times  \\
&& \frac{ \Gamma^{-\alpha-1} E^{-\alpha} (1-\beta  \cos(\theta) \sqrt{1-m^2 c^4/E^2})^{-\alpha-1} }
{ \left[ \sin^2 (\theta) + \Gamma^2 \left(  \cos(\theta) - \frac{ \beta  }{ \sqrt {1 - m^2c^4 / E^2} } \right)^2 \right]^{1/2} }
\nonumber
\end{eqnarray}

Note that in the limit, when $m c^2\ll E$, the latter expression reduces to 
$ (A/4\pi)  E^{-\alpha}  D^{2+\alpha}$.
To prove the latter we
use that
$
\left(\sin^2 (\theta) +\Gamma^2 (\cos(\theta)-\beta)^2\right)^{1/2} = 1/D ,   
$           
which can be established by squaring both sides, and 
summing and subtracting $\Gamma^2 \sin^2(\theta) + \Gamma^2 \beta^2 \cos^2 (\theta)$
on the left hand side, with the additional use of the definition of $\Gamma^2 = 1 / [ 1- \beta^2]$.
Note also 
that we are not assuming
a specific jet composition, thus it is a general result for any kind of
particle distributions.
 Figure \ref{n-fig} shows the resulting expression (\ref{final}) for a power-law spectrum of protons in a beam as seen by the observer, for different values of viewing angles, and an example of Lorentz factor and slope (solid lines). Note that for high energies, when $E>\Gamma mc^2$, $n(E,\Omega$) 
 is a power-law distribution, as the original primed one, albeit it presents a  different behavior for lower values of $E$.

\subsection{A power-law with an exponential  cutoff}
\label{expc}

We shall briefly also consider the case in which the intrinsic particle distribution in the beam is a 
power-law with an exponential cutoff. This would be a natural resulting consequence of particles being subject to losses on the same 
site where they are accelerated. As such, the cutoff appears in the primed referenced frame, where
\be
n'(E',\Omega') = (A/4\pi)  E'^{-\alpha} \exp{(-E'/E_{\rm cut})}.       
\label{n'cut}                       
\ee
Note that $E_{\rm cut}$ should be fixed in the beam frame, where acceleration and losses are supposed to occur, but once fixed it is at all effects a number, and it is not subject to Lorentz transformations. However, the energy variable in the exponential function is. The result  is
\begin{eqnarray} 
\label{final-cut}
n(E,\Omega)  \hspace{-.2cm} &&=  \frac{ A}{4\pi} \times  \\
&& \frac{ \Gamma^{-\alpha-1} E^{-\alpha} (1-\beta  \cos(\theta) \sqrt{1-m^2 c^4/E^2})^{-\alpha-1} }
{ \left[ \sin^2 (\theta) + \Gamma^2 \left(  \cos(\theta) - \frac{ \beta  }{ \sqrt {1 - m^2c^4 / E^2} } \right)^2 \right]^{1/2} } \times \nonumber \\
&& \exp \left(  -\frac{\Gamma E } {E_{\rm cut}} (1-\beta \cos(\theta) \sqrt{1- m^2c^4/E^2}
\right).
\nonumber
\end{eqnarray} 
When $mc^2 \ll E$, the exponential cutoff $\exp{(-E'/E_{\rm cut})}$ gets transformed into
$\exp{(-E/DE_{\rm cut})}$, with $D$ being the bulk Doppler factor.

\subsection{Comparison with Purmohammad \& Samimi (2001) }
\label{CT}

\begin{figure*}
\centering
\vspace{0.4cm}
\includegraphics[angle=-90, scale=0.3]{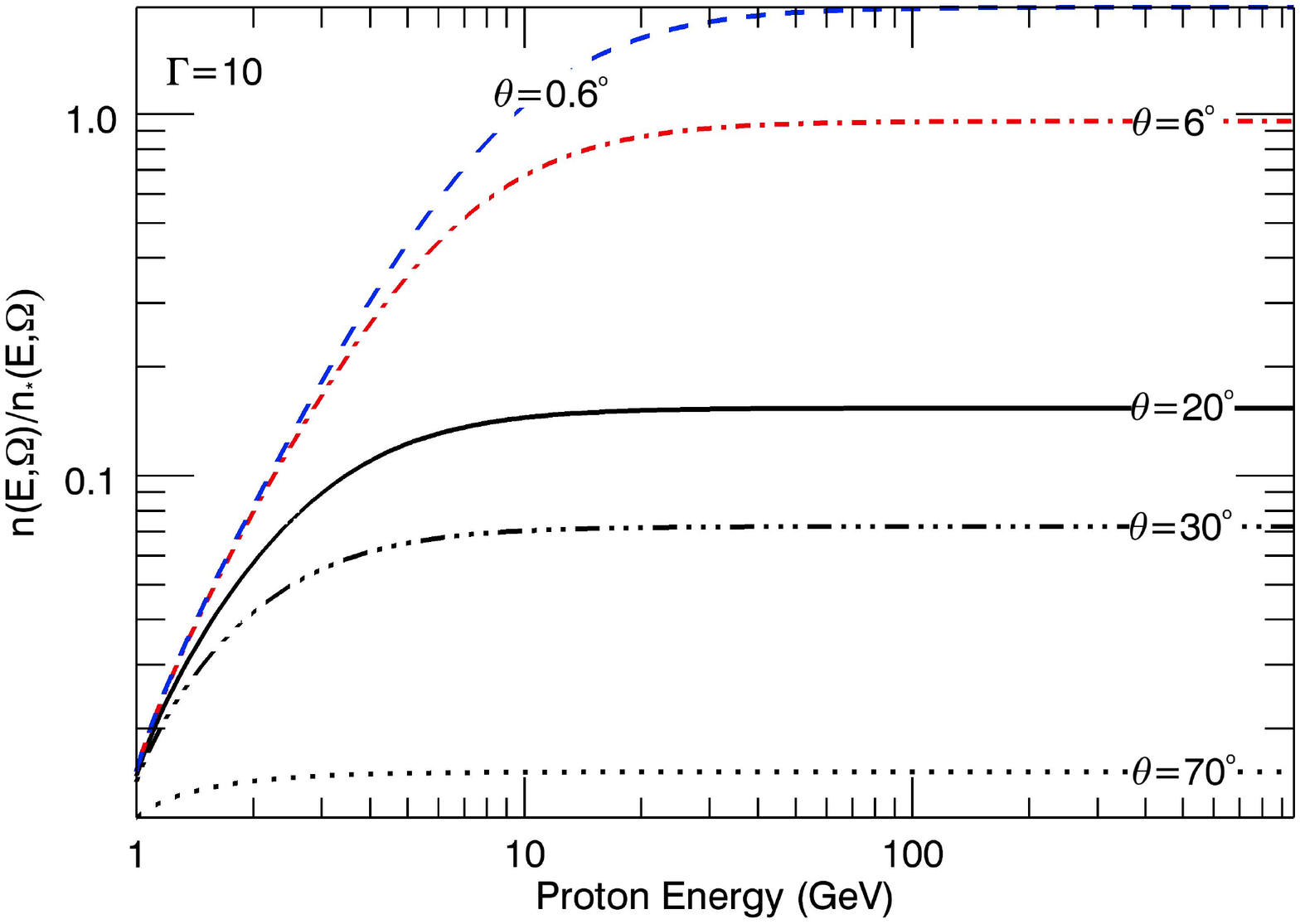} \hspace{0.5cm}
  \includegraphics[angle=-90, scale=0.3]{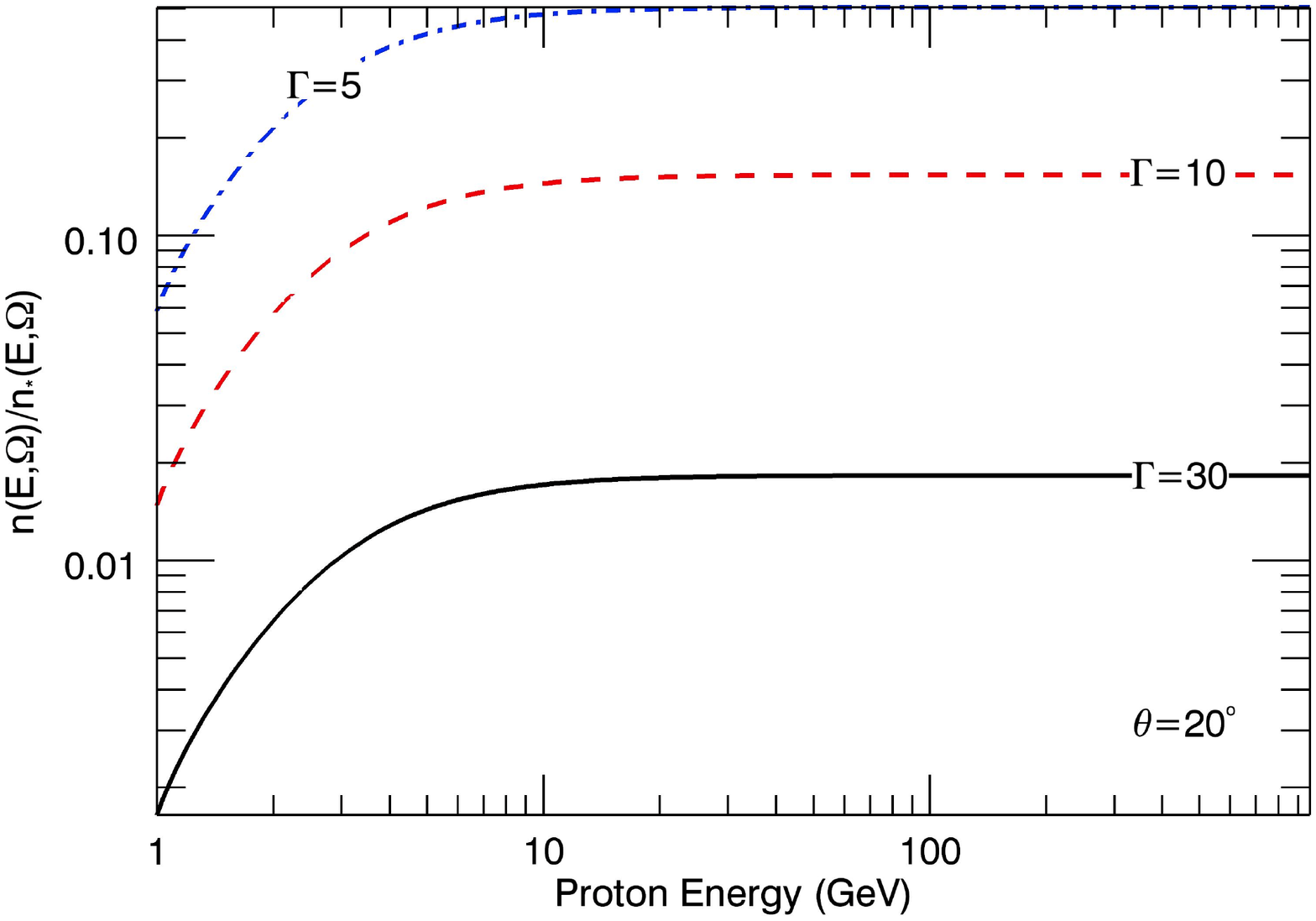}
     \caption{Comparison of the derived expression for the spectrum of protons in beams, following Equation (\ref{comp}).
 Left: as a function of the viewing angle ($\theta$) for $\Gamma=10$. Right: as a function of the bulk Lorentz factor ($\Gamma$), for $\theta=20^o$.
}
         \label{image}
\end{figure*}

Purmohammad and Samimi  (PS01) states that the isotropic particle distribution transformed from the
jet into the observer frame results in (in this subsection and to avoid confusion we label their expressions with a $\star$):
   \begin{eqnarray} 
\label{them-general}
n_\star(E,\Omega) &= &\frac{A}{4\pi} 
 \Gamma^{-\alpha+1}  E^{-\alpha}  \times
\\
&&  \frac {(1-\beta \cos(\theta) \sqrt{1-m^2 c^4/E^2} )^{-\alpha} }
{ \left[ \sin^2 \theta +\Gamma^2 \left(\cos(\theta)- \frac{\beta }{ \sqrt{1-m^2 c^4/E^2}}\right)^2 \right]^{1/2} }. \nonumber 
\end{eqnarray}    
Purmohammad and Samimi's (2001) paper does not provide a derivation of their Eq. (3), quoted above as (\ref{them-general}), but earlier private communication with one of us was sufficient to reconstruct it.
The steps are the following. They start by assuming conservation of particles in the form
\be
n_\star(E,\Omega)dVd\Omega dE = n_\star'(E', \Omega')  dV'   d\Omega'  dE',	
\label{prob}
\ee
thus one has
\be
n_\star(E,\Omega)= n_\star'(E'(E, \theta), \Omega'(E, \theta)) \frac{dV' } {dV}  \frac{ d\Omega'} {d\Omega}  \frac{dE'	 } {dE}.
\label{key}
\ee
From the Lorentz-invariance of the phase-space element $d {\bf p} / E$, 
the invariance of $p d\Omega dE$ can also be proved. 
Then, one can write that 
\be
\frac{p}{p'} = \frac{dE'} {dE}  \frac{d\Omega'} {d\Omega} .
\label{key2}
\ee
Using the expressions derived above for $p/p'$, and the equality $dV'=\Gamma dV$,  into
(\ref{key}), one gets 
 \be
n_\star(E,\Omega)= \frac{ \Gamma (A/4\pi)  E'^{-\alpha} }{
\sqrt{\sin^2 (\theta) + \Gamma^2 (  \cos(\theta) - \beta E / \sqrt {E^2 - m^2c^4} )^2 } }.
 \ee
 From here, after replacing in the numerator
 $ E' = \Gamma ( E- \beta c p \cos (\theta) ) 
        =  \Gamma ( E - \beta  \cos (\theta) \sqrt{E^2 - m^2 c^4} ) $,
the expression (\ref{them-general}) immediately
follows.


Formally, one can say that there is a problem hidden in Eq. 
(\ref{prob}), from where the rest of the derivation follows. The invariant phase volume is not $dV d\Omega dE$ but rather $p^2 dV d\Omega dp $. To avoid 
inconsistencies, invariants (from where
to derive transformation laws) should be written as operations over individually invariant factors, e.g., $dN$ over $d\nu$, rather than from identities
(which is what one obtains replacing the definition of $n$ and $n'$ in Eq. (\ref{prob})). However, the latter can also yield to the correct result too if care is exercised. 
In practice, then, the problem with the prior derivation is in the use of the equality $dV'=\Gamma dV$. The latter equation implies that $dV'$ is the proper volume $dV_0$; which is clearly incorrect, given the fact that the momentum, ${\bf p'}$, of the particles in that volume is not zero. Only for the proper volume the referred equality is true (see, e.g., eq. 4.6 of Landau \& Lifshitz 1987). To clarify this issue, let us introduce, in addition of the two reference systems (observer and beam), another frame $K_0$ in which the particles with the given momentum are at rest. The proper volume $dV_0$ of the element occupied by the particles is defined relative to this system. The velocities of the primed and unprimed systems relative to the system $K_0$ coincide by definition with the velocities ${\rm v}$ and ${\rm v'}$ which these particles have in the systems of the observer and the beam, respectively. Thus, we can write
\be
dV= dV_0 \sqrt{ 1 - {\rm v}^2 / c^2}, \hspace{0.2cm}
dV'= dV_0 \sqrt{ 1 - {\rm v'}^2 / c^2}
\ee
from which one can write
\be
\frac{dV}{dV'} =  \frac {\sqrt{ 1 - {\rm v}^2 / c^2}} {\sqrt{ 1 - {\rm v'}^2 / c^2}} = \frac{E'}{E}.
\label{VV}
\ee
If we use Eqs. (\ref{key2}) and (\ref{VV})  in Eq. 
(\ref{prob}), the latter transforms back into the expression of the invariant quantity ${1}/(p E) \times {dN}/({dV dE d\Omega}) $ and Eq. (\ref{gen-inv}) follows.

The difference between expression (\ref{them-general})
and our Eq. (\ref{final}) is important. Diving it one by another we obtain
\be
\frac{n(E,\Omega)  } { n_\star(E,\Omega) } = 
\Gamma^{-2} (1-\beta  \cos(\theta) \sqrt{1-m^2 c^4/E^2})^{-1} 
\label{comp}
\ee
which is a  non-trivially dependent function of $\Gamma$, $E$, and $\theta$. However, note that it is not a function of $\alpha$.
In the extremely relativistic case; $m c^2\ll E$ 
one notices a factor $D / \Gamma$ of difference. Put otherwise, in the stated asymptotic regime, the ratio expressed by Eq. (\ref{comp}) reduces to 
$
({n(E,\Omega)  } /  { n_\star(E,\Omega) } ) = 
\Gamma^{-2} (1-  \cos(\theta) \sqrt{1-1/\Gamma^2})^{-1} ,
$
which is only a function of $\Gamma$ and $\theta$. 
One can see that in the case of $\theta = 0^o$, 
${n(E,\Omega)  } / { n_\star(E,\Omega) } 
\rightarrow 2$ for growing $\Gamma$. For other values of $\theta$, this 2D function
quickly fall off.
We plot the comparison between our derivation 
and that by PS01 in Figure \ref{image}, both as a function of 
viewing angle $\theta$ (for fixed $\Gamma=10$) and bulk Lorentz factor $\Gamma$ (for $\theta=20^o$). We see that the differences amounts from a factor of $\sim 2$
in the case of directly-pointing beams of blazars and microblazars (of very low $\theta$, in our example, 0.6$^o$) to one order of magnitude at least for the more common cases of moderate
inclinations (e.g., for 20$^o$; with the difference being significantly larger for larger inclinations). Note that as soon as $\theta$ is larger than $\sim 5^o$,
the proton spectrum in the beam increasingly undershoots  PS01's calculation; this leads to an analogous 
correction of previously calculated $\gamma$-ray and neutrino emission (typically by more than one order of magnitude) in all cases in which the expression (\ref{them-general})
was used. Note also that in
  the low energy decline PS01's calculation always overestimates significantly (by about two orders of magnitude) the correct yield of the proton spectrum (see Fig. \ref{n-fig} at proton energies between 1 and 10 GeV).

\section{Discussion}

The proton beam model presented by PS01 gives an expression for a power-law proton spectrum 
in a beam, as seen by the observer. We have demonstrated that this expression 
is incorrect, since it is not  complying with relativistic invariants. 
We have derived the correct transformation 
from the jet frame particle density to the 
observer's, which strictly obeys
 these Lorentz invariance constraints. By comparing our result 
with PS01, we noticed that differences are strongly dependent on the viewing angle. For many cases,
PS01 calculation entails a significant over-prediction (order of magnitude) 
of the proton spectra for $E>\Gamma
mc^2$, whereas it always significantly over-predicts (two orders of magnitude) the proton spectrum 
at lower energies, disregarding the viewing angle.
The $\gamma$-ray luminosity is related to the particle density through the corresponding emissivity of the processes by which the particles interact. 
For instance, the charged and neutral pion decay channels is the most effective yield of photons and neutrinos in many situations, and its emissivity is directly proportional
to the particle density $n(E,\Omega$). 
A large number of publications relying on the calculation by Purmohammad \&
Samimi (2001) for estimating $\gamma$-ray and neutrino emission yields in hadronic
beam scenarios are thus affected in the way described above.

One particular example can be seen in the hadronic MQ case (Romero et al. 2003)\footnote{
Note that the eq. (2) in the latter paper, 
has a missing square in the denominator factor $(\cos(\theta)-\beta E/ \sqrt(E^2-m^2 c^4))$. This typo has been adopted in several of the
papers quoting Romero et al (2003). 
}
 {and their applications (e.g., Romero et al. 2005, Orellana et al. 2007, Reynoso et al. 2008).
These are all affected on the light of our results. MQs can have  
$\Gamma \sim 2$, but $\theta$-values above 10$^0$ are typical. MQs can precess, as is the case of SS433, where  $\theta$ can change even up to 20$^0$ around an already large mean viewing angle. }

Only a small part of the $\gamma$-ray AGNs belong to the non-blazar class. The majority among those associated with radio sources possess relatively large core dominance (Abdo et al. 2010), indicating that their viewing angle is likely not large. Assuming  $\theta <20^o$, if $\Gamma < 10$, 
PS01 deviation is not more than a factor of $\sim 6$
(at $E>mc^2$). For the {\it average} blazar with $\Gamma =10$ and $\theta <10^o$ the difference is a factor of $\sim 2$. However, it has been argued that during extreme TeV flaring events of some blazars (e.g. Mkn 501, PKS 2155-304), Lorentz factors greater than $\Gamma \sim 50$ are required (e.g. Begelman et al. 2008). If the viewing angle remains at $\theta \sim 1/\Gamma_{\rm quiet} \sim 1/10 \sim 5.7^o$ during such events, 
PS01 formulation error 
can grow to more than an order of magnitude. 
Furthermore, a few radio galaxies with large viewing angles are also detected at $\gamma$-ray energies: M87 with viewing angle $\theta < \sim40^o$, and Cen A where estimates for the viewing angle range from $15^o$ to $80^o$. These radio galaxies have been already modeled with the PS01 derivation   (e.g., Reynoso et al. 2010). Here, the difference between the PS01 formula and ours can reach up to a factor 20 for $\Gamma$ not larger than 5.

Finally, we note that models designed to predict future detections of large-viewing angle sources with higher sensitivity instruments than currently available will greatly overpredict the number of sources if the PS01 formula is used. Similarly, a model for an  AGN which overestimates the $\gamma$-ray flux could make wrong associations of unidentified $\gamma$-ray sources.
Given that  $\gamma$-ray fluxes will be in many cased diminished, and that the time of observations in Cherenkov facilities 
grows with the square of the flux-reduction factor, the likelihood of observing hadronic beams is undermined.


\acknowledgements

DFT acknowledges support from the grants AYA2009-07391, SGR2009-811, and
TW2010005.
AR acknowledges support by the Marie Curie IRG grant 248037 within the FP7 Program.
We acknowledge O. Reimer for discussions.

\end{document}